# Principal Component Analysis on the Philippine Health Data

**M. Carillo[1], F. Largo[2] and R. Ceballos[3]**

Department of Mathematics and Statistics, College of Arts and Sciences
University of Southeastern Philippines, Davao City, Philippines

Email : mcarillo211995@gmail.com[1]
fe1974@yahoo.com[2]
roel.ceballos@usep.edu.ph[3]

**ABSTRACT**

*This study was conducted to determine the structures of a set of n correlated variables and creates a new set of uncorrelated indices which are the underlying components of the Philippine health data. The data utilized in this study was the 2009 Philippine Health Data which was made available by National Statistical Coordination Board(NSCB) in its 2009 publication. The publication contains the health data of 81 provinces of the Philippines consisting of ten system-related determinants which was considered as the variables in this study. From the ten health system-related determinants, it was found out that there are three significant underlying components that could summarize the Philippine health data. The first component was named as "importance of safe water supply and emphasis on child heat" while the second and third component were named as "importance of Barangay Health Stations, government health workers and emphasis on pregnant women's health" and "emphasis on women's health", respectively. These three components jointly account for a total of 73.01 % of the total variance explained by the component.*

## 1. INTRODUCTION

Health has a number of determinants. Difficulty in examining the interrelationship among these determinants is unquestionable and it is difficult to interpret data with higher dimensions. Health data tends to be multidimensional. Hence, it is important to use dimension reduction especially in health data for it will provide a clearer picture and visual examination of data of interest. Through this, health care providers and other stakeholders in the healthcare delivery system can develop more thorough and insightful diagnosis and treatments, resulting to a higher quality of health.

Given a set of data with a large number of variables, principal components analysis is a technique that can be used to reduce the variables into fewer dimensions created as a linear combination of the original number of variables. PCA is a factor analysis technique that will identify a smaller number of uncorrelated variables called principal component that will account for the variation in the observed variables (Jollife, 2002). With this, the researcher was impelled to study on the Philippine Health Data using Principal Component Analysis (PCA), for which the researcher will determine the number of underlying components that summarizes the Philippine Health Data. The aims are to analyze the Philippine health data using Principal component analysis and to determine the number of components that summarizes the health data of the Philippines.

## 2. METHODS

Principal Component Analysis (PCA) was used as a data reduction tool. It employed the use of varimax rotation as the type of orthogonal rotation.

Kaiser-Meyer Olkin. It refers to the calculated measures of the entire correlation matrix and each individual variable which evaluates the appropriateness of applying PCA.

Scree Plot. Scree plot displays the eigenvalues associated with the component or factor in descending order versus the number of the component. In PCA, it is used to visually asses which components most of the variability in the data (Jollife, 2002).

Measure of Sampling Adequacy (MSA). It measures if the correlations between xi and other variables are unique, that is not related to the remaining variables outside its simple correlation. Kaiser as cited by (Wuensch, 2012) has described MSAs above 0.90 as marvellous and below 0.50 as unacceptable.

## 3. RESULTS
### 3.1. Preliminary Analysis

The analysis of Table 1 shows the evaluation of the data assumptions for Principal Component Analysis. It includes the value of Kaiser-Meyer Olkin(KMO) which is used to determine if the Principal Component Analysis is useful for these variables. Based on the result, the KMO value is equal to 0.645 which suggest that data is appropriate for applying PCA.

*Table 1:* Evaluation of the Data

| KMO | Evaluation |
|---|---|
| 0.645 | Satisfied. PCA is appropriate for the analysis of these variables |

Bartlett's test on the other hand is used to test if interrelationship among variables is present. It is shown in Table 2 that Bartlett's test of sphericity has a p-value which is less than 0.01; hence, the requirement for the interrelationship among variables is satisfied in this study.

*Table 2:* Evaluation for Interrelationship

| Batlett's Test | p-value | Evaluation |
|---|---|---|
| 340.002 | 0.000* | Satisfied. Variables are interrelated |

*actual p-value is less than 0.01

One of the ways to check whether the variables are significant for the analysis of principal component analysis is to check its Measure of Sampling Adequacy (MSA). It is suggested that an MSA value of less than 0.50 (unacceptable) must be excluded for further analysis because the correlations between Xi and the other variables are unique, that is not related to the remaining variables outside each simple correlation. Table 3 displays the Anti-Image Correlation Matrix. On the main diagonal of the matrix are the overall MSA's of the individual variables.

Table 3: Evaluation for Interrelationship

|     | $X_1$ | $X_2$ | $X_3$ | $X_4$ | $X_5$ | $X_6$ | $X_7$ | $X_8$ | $X_9$ | $X_{10}$ |
|-----|-------|-------|-------|-------|-------|-------|-------|-------|-------|----------|
| $X_1$ | .671a | -.060 | .022 | .041 | .109 | -.007 | -.123 | .222 | .141 | -.377 |
| $X_2$ | -.060 | .635a | .186 | -.145 | -.198 | .268 | -.068 | .048 | -.863 | -.059 |
| $X_3$ | .022 | .186 | .739a | -.629 | -.095 | .011 | -.039 | .118 | -.293 | -.216 |
| $X_4$ | .041 | -.145 | -.629 | .737a | -.166 | .147 | -.153 | -.322 | .025 | .205 |
| $X_5$ | .109 | -.198 | -.095 | -.166 | .693a | -.314 | .141 | .063 | .196 | .010 |
| $X_6$ | -.007 | .268 | .011 | .147 | -.314 | .499a | -.765 | -.174 | -.230 | .072 |
| $X_7$ | -.123 | -.068 | -.039 | -.153 | .141 | -.765 | .596a | .151 | -.002 | -.068 |
| $X_8$ | .222 | .048 | .118 | -.322 | .063 | -.174 | .151 | .587a | -.044 | -.292 |
| $X_9$ | .141 | -.863 | -.293 | .025 | .196 | -.230 | -.002 | -.044 | .668a | .036 |
| $X_{10}$ | -.377 | -.059 | -.216 | .205 | .010 | .072 | -.068 | -.292 | .036 | .335a |

Notice that the MSA values for variables $X_6$ and $X_{10}$ are fairly low with a value of 0.499 and 0.335, respectively. This suggests that these variables are excluded for the final analysis of the data since its MSA value is less than 0.50 which is described by Kaiser as unacceptable and the correlations between these variables and the other is not unique. Hence, these variables might not be explained well by the identified components and must be removed for further analysis.

### 3.2. Principal Component Analysis

After variables $X_6$ and $X_{10}$ have been removed from the analysis, PCA procedure is conducted. At this point, only eight variables were included in the analysis namely: $X_1$, $X_2$, $X_3$, $X_4$, $X_5$, $X_7$, $X_8$, and $X_9$. With only eight variables included in the analysis, the KMO (see Table 4) still satisfies the appropriateness of conducting PCA with a value of 0.709. Furthermore, as shown in Table 5, Bartlett's test of sphericity has a p-value which is less than 0.01 thus suggesting that the assumption for interrelation of variables is satisfied.

Table 4: Evaluation of Data in the Second Analysis

| KMO | Evaluation |
|-----|------------|
| 0.709 | Satisfied. PCA is appropriate for the analysis of these variables |

Notice that the Bartlett's test of sphericity as shown in Table 5 has a p-value which is less than 0.01 thus suggesting that the assumption for interrelation of variables is satisfied.

Table 5: Interrelationship of Variables in the Second Analysis

| Batlett's Test | p-value | Evaluation |
|----------------|---------|------------|
| 254.416 | 0.000* | Satisfied. Variables are interrelated |

*actual p-value is less than 0.01

Table 6 shows the new MSA values of the final variable included in the analysis. It can be observed that the MSA values for the eight variables may not be marvellous, but they are not low enough to be dropped for the final analysis of the data.

*Table 6:* Anti-Image Correlation Matrix

|    | X1 | X2 | X3 | X4 | X5 | X7 | X8 | X9 |
|----|------|------|------|------|------|------|------|------|
| X1 | .756[a] | -.078 | -.084 | .132 | .150 | -.180 | .124 | .156 |
| X2 | -.078 | .642[a] | .173 | -.199 | -.096 | .174 | .143 | -.858 |
| X3 | -.084 | .173 | .752[a] | -.615 | -.094 | -.104 | .041 | -.305 |
| X4 | .132 | -.199 | -.615 | .775[a] | -.081 | -.090 | -.216 | .064 |
| X5 | .150 | -.096 | -.094 | -.081 | .767[a] | -.181 | -.221 | .130 |
| X7 | -.180 | .174 | -.104 | -.090 | -.181 | .644[a] | .120 | -.175 |
| X8 | .124 | .143 | .041 | -.216 | -.221 | .120 | .745[a] | -.129 |
| X9 | .156 | -.858 | -.305 | .064 | .130 | -.175 | -.129 | .660[a] |

In deciding how many components are to be retained, the total eigenvalues must be inspected. As in the Kaiser's rule, component with total eigenvalue of greater than 1 will be retained. Components 1, 2 and 3 have total eigenvalues of 3.520, 1.209 and 1.112, respectively (see Table 7). Since their calculated eigenvalues are greater than 1, they are the only components to be retained.

*Table 7:* Anti-Image Correlation Matrix

| Component | Total Eigenvalues |
|-----------|-------------------|
| 1 | 3.520 |
| 2 | 1.209 |
| 3 | 1.112 |
| 4 | 0.710 |
| 5 | 0.613 |
| 6 | 0.540 |
| 7 | 0.216 |
| 8 | 0.081 |

Another way of identifying the meaningful component to be retained is by looking at the scree plot which is a graph of the eigenvalues associated with each component. The rule is to look for a "break" between the components separating large eigenvalues from small eigenvalues. From the scree plot shown in Figure 15, it can be seen that component 3 is the breaking point in the plot since the eigenvalues begin to level off and is considered as the last component to be retained.

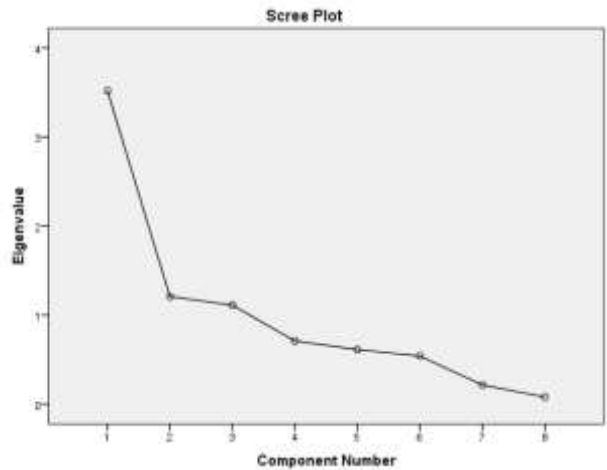

**Figure 1.** Scree Plot

Table 8 shows the cummulative percentage of the total variance explained by the three components. It is shown that the first component accounts for 43.996 percent of the total variance. The second component accounts for 15.113 percent of the total variance that is not accounted for by the first component. Furthermore, a total of 13.902 percent is accounted by the third component and is not accounted for by the first and second component. These three factors jointly accounts for 73.010 percent of the total variance explained by the health data. That is to say that the variance of the standardized health data with the eight variables has been well accounted for by these identified components.

*Table 8:* Percentage of Total Variance Explained by Components

| Component | Total Eigenvalues | Percentage of Variance | Cumulative Percentage |
|---|---|---|---|
| 1 | 3.520 | 43.996 | 43.996 |
| 2 | 1.209 | 15.113 | 59.109 |
| 3 | 1.112 | 13.902 | 73.01 |

Component loadings for the three components are shown in Table 9. A component loading greater than 0.50 regardless of the sign is highlighted. Notice that the structure of the components in the table is complex. Meaning, there is at least one variable that load heavily on more than one component and it is more important to have a simple component structure where variables load heavily on component and load lowly on the other component. Since variable X7 is loaded heavily on more than one component, rotation must be done. Varimax rotation is used since it tends to maximize the variance of the component pattern matrix.

*Table 9:* Component Loadings of the Eight Variables

|  | Component | | |
|---|---|---|---|
|  | 1 | 2 | 3 |
| $x_1$ | -0.416 | 0.575 | 0.302 |
| $x_2$ | 0.799 | 0.267 | -0.386 |
| $x_3$ | 0.832 | 0.154 | 0.123 |
| $x_4$ | 0.861 | 0.016 | 0.080 |
| $x_5$ | 0.465 | -0.417 | 0.514 |
| $x_7$ | 0.342 | 0.514 | 0.685 |
| $x_8$ | 0.461 | -0.610 | 0.129 |
| $x_9$ | 0.854 | 0.255 | -0.317 |

Table 10 shows the rotated component loadings sorted by the size of its corresponding loadings after a varimax rotation is done with Kaiser Normalization. It can be seen that after a component rotation has been done, the component structure in Table 10 is more simple and clean than of Table 9 since each of the variables load heavily only on one component, hence making the pattern easy to interpret.

*Table 10:* Component Loadings of the Eight Variables

|  | Component | | |
|---|---|---|---|
|  | 1 | 2 | 3 |
| $x_1$ | -0.273 | -0.564 | 0.450 |
| $x_2$ | 0.924 | 0.028 | -0.059 |
| $x_3$ | 0.725 | 0.307 | 0.334 |
| $x_4$ | 0.718 | 0.419 | 0.239 |
| $x_5$ | 0.066 | 0.732 | 0.339 |
| $x_7$ | 0.183 | 0.053 | 0.849 |
| $x_8$ | 0.139 | 0.758 | -0.085 |
| $x_9$ | 0.942 | 0.087 | 0.005 |

There are four variables which are loaded heavily on the first component namely: number of fully immunized children (0-9 months) (x2); number of children (12-59 months) given Vitamin A (x3); number of children (6-11 months) given Vitamin A (x4) and number of household with access to safe water supply (x9). Thus, component 1 measures the extent, to which provinces receive and restore health services by immunization of children, care for children by providing Vitamin A and having access to safe water supply. It can be classified as "importance of safe water and emphasis on child health". On the second component, two variables have a high positive loading and one with a negative high loading namely: number of barangay health workers (x1); percentage of pregnant women given TT2 (two doses of Tetanus Toxoid) (x5); and ratio of Barangay Health Stations (BHS) (x8). Thus, component 2 measures the extent to which provinces receive and restore health services by care for pregnant women and having enough number of BHS but lack of government health workers. It can be classified as "importance of Barangay Health Station, government health workers and emphasis on pregnant women's health". The third components have only one variable with high loading namely percentage of pregnant women given complete Iron ($x_7$). Thus, the third component measures the extent to which provinces receive and restore health services by care for women. It can be classified as "emphasis on women's health."

From the rotated component loadings in Table 10, the principal component model can be obtained from dividing the component loading by the square root of the corresponding eigenvalue. The following are the produced model for estimating the component scores from the data values (after the variables have been standardized to have a mean zero and unit standard deviations):

$PC_1$ = -0.145$x_1$ + 0.492$x_2$ + 0.386$x_3$ + 0.382$x_4$ + 0.035$x_5$ + 0.098$x_7$ +0.074$x_8$ + 0.502$x_9$

$PC_2$ = -0.513$x_1$ + 0.025$x_2$ + 0.279$x_3$ + 0.381$x_4$ + 0.666$x_5$ + 0.048$x_7$ +0.689$x_8$ + 0.079$x_9$

$PC_3$ = 0.427$x_1$ - 0.056$x_2$ + 0.317$x_3$ + 0.227$x_4$ + 0.321$x_5$ + 0.805$x_7$ -0.080$x_8$ + 0.005$x_9$

## 4. DISCUSSION AND CONCLUSION

The researcher concluded that there were three components identified from the health data of the Philippines. Component 1 is labelled as "importance of safe water and emphasis on child health"; Component 2 as "importance of Barangay Health Stations (BHS), government health workers and emphasis on pregnant women's health" and Component 3 as "emphasis on women's health". The researcher recommended that future researcher should explore more on the other field of research in which the Principal Component Analysis can be applied as a reduction technique.